\documentclass[twocolumn]{aastex62}

\received{}
\revised{}
\accepted{}
\submitjournal{AJ}

\shorttitle{Confirmation of WASP-107b's extended Helium atmosphere with Keck II/NIRSPEC}
\shortauthors{Kirk et al.}

\begin{document}

\title{Confirmation of WASP-107b's extended Helium atmosphere with Keck II/NIRSPEC}

\correspondingauthor{James Kirk}
\email{james.kirk@cfa.harvard.edu}

\author[0000-0002-4207-6615]{James Kirk}
\affil{Center for Astrophysics $\vert$ Harvard \& Smithsonian, 60 Garden Street, Cambridge, MA 02138, USA}

\author[0000-0003-4157-832X]{Munazza K.\ Alam}
\affil{Center for Astrophysics $\vert$ Harvard \& Smithsonian, 60 Garden Street, Cambridge, MA 02138, USA}

\author[0000-0003-3204-8183]{Mercedes L\'{o}pez-Morales}
\affil{Center for Astrophysics $\vert$ Harvard \& Smithsonian, 60 Garden Street, Cambridge, MA 02138, USA}

\author[0000-0003-1957-6635]{Li Zeng}
\affil{Department of Earth and Planetary Sciences, Harvard University, Cambridge, MA 02138, USA}



\begin{abstract}

We present the detection of helium in the extended atmosphere of the sub-Saturn WASP-107b using high resolution ($R \approx 25000$) near-infrared spectra from Keck II/NIRSPEC. We find peak excess absorption of $7.26 \pm 0.24\%$ (30\,$\sigma$) centered on the He\,I triplet at 10833\,\AA. The amplitude and shape of the helium absorption profile is in excellent agreement with previous observations of escaping helium from this planet made by CARMENES and HST. This suggests there is no significant temporal variation in the signature of escaping helium from the planet over a two year baseline. This result demonstrates Keck II/NIRSPEC's ability to detect atmospheric escape in exoplanets, making it a useful instrument to further our understanding of the evaporation of exoplanetary atmospheres via ground-based observations of He\,I. 

\end{abstract}

\keywords{methods: observational -- planets and satellites: atmospheres, gaseous planets, individual (WASP-107b)}


\section{Introduction} \label{sec:intro}

Recently, \cite{Spake2018} announced the ground-breaking detection of He\,I in the extended atmosphere of the sub-Saturn exoplanet WASP-107b, using the G102 grism on Hubble Space Telescope's (HST) Wide Field Camera 3 (WFC3) instrument. \cite{Spake2018} detected an excess absorption of $0.049\pm0.011 \%$ in a 98\,\AA -wide bin centered on the helium triplet at 10833\,\AA. This absorption suggested that the planet is losing mass at a rate of $10^{10} - 3 \times 10^{11}$\,g\,s$^{-1}$, or equivalently $0.1 - 4$\,\% of its total mass per Gyr \citep{Spake2018}.

Prior to the detection by \cite{Spake2018}, this absorption signature of a metastable state of neutral helium at 10833\,\AA\ had been predicted to be a strong feature in exoplanet transmission spectra \citep{Seager2000,Turner2016,Oklopcic2018}. However, despite early efforts \citep{Moutou2003}, it went undetected until \cite{Spake2018}'s detection. 

The result of \cite{Spake2018} coincided with a theoretical study of \cite{Oklopcic2018}, who generated 1D models of escaping atmospheres and demonstrated that observations of He\,I absorption could be used as a powerful tool for studying atmospheric escape. This is because it suffers little from extinction, can be observed from the ground, and can probe the escaping planetary wind. 

Prior to the detection of helium in WASP-107b's atmosphere, the detections of extended atmospheres had primarily been achieved via studies of Lyman-$\alpha$ at UV wavelengths (e.g.\ \citealp{Vidal-Madjar2003,Lecavelier2010,Ehrenreich2015,Bourrier2018}), which can only be observed with HST. These studies have revealed enormous clouds of escaping hydrogen, with absorption depths as large as 56.3\,\%  (GJ\,436b, \citealt{Ehrenreich2015}), which have enabled detailed studies of the cloud properties and dynamics (e.g.\ \citealp{Bourrier2013,Bourrier2016}). However, Lyman-$\alpha$ is strongly affected by both interstellar extinction, limiting observations to planets within $\sim 50$\,pc (e.g.\ \citealp{Jensen2018}), and contamination from geo-coronal emission (e.g.\ \citealp{Ehrenreich2015}).

In addition to studies of Lyman-$\alpha$, H-$\alpha$ has also been used as a probe of extended exoplanet atmospheres, with a handful of detections to date (\citealp{Casasayas-Barris2018,Jensen2012,Jensen2018,Yan2018}). However, this line is more strongly affected by the host star's activity than the metastable helium triplet, where stellar activity is more likely to dilute the signal than to amplify it, making misinterpretation regarding the planetary nature of any absorption less likely \citep{Cauley2018}. This makes the metastable helium triplet attractive for studies of extended atmospheres, even in the case of active host stars \citep{Cauley2018}.

Studies of the helium triplet can therefore provide us with a larger sample of evaporating exoplanets than is currently achievable via observations of Lyman-$\alpha$ and H-$\alpha$. This will allow for further constraints to be placed on models of escaping exoplanet atmospheres \citep{Oklopcic2018}. By building the sample of exoplanets known to be undergoing atmospheric loss, we can further constrain the role that photoevaporation plays in shaping the population of observed exoplanets. Evaporation is thought to be responsible for the `Neptune desert', a region of parameter space where there is a dearth of Neptunian exoplanets  (e.g.\ \citealp{Mazeh2016}), and the gap in the radius distribution of small planets between 1.5 and 2 Earth radii \citep{Fulton2017,Owen2017,Zeng2017}. The ability to observe evaporation from the ground will allow for robust tests of the role of evaporation in shaping exoplanet populations.

Indeed, there have been six ground-based detections of extended helium atmospheres since the detection of \cite{Spake2018}, with one additional space-based detection \citep{Allart2018,Allart2019,Mansfield2018,Nortmann2018,Salz2018,Alonso-Floriano2019,Ninan2019}. However, there have also been four non-detections of extended helium atmospheres \citep{Kreidberg2018_w12,Nortmann2018,Crossfield2019}, with the stellar XUV flux likely playing a key role \citep{Nortmann2018,Oklopcic2019}.

Here, we present the detection of the extended helium atmosphere of WASP-107b using the high resolution ($R = 25000$) near-infrared spectrograph NIRSPEC on the Keck II telescope. This is the first time this instrument has been used to detect helium in an exoplanet's atmosphere.

WASP-107b, detected by \cite{Anderson2017}, is a warm sub-Saturn with a mass of 0.119\,M$_\mathrm{J}$, radius of 0.924\,R$_\mathrm{J}$, and equilibrium temperature of 736\,K \citep{Mocnik2017}. Because of its deep transit (2.09\,\%, \citealp{Mocnik2017}) and large atmospheric scale height (855\,km, derived from \citealp{Mocnik2017}), it is an excellent target for atmospheric studies. In addition to the detection of escaping He I \citep{Spake2018,Allart2019}, \cite{Kreidberg2018_w107} detected water using HST/WFC3's G141 grism and found evidence for a methane-depleted atmosphere and high altitude condensates.

\section{Observations} \label{sec:obs}

We observed a single transit of the planet WASP-107b on the night of April 6th 2019 using the NIRSPEC near-infrared spectrograph on the Keck II telescope at Mauna Kea Observatory, Hawai'i. We acquired 36 spectra of the target over 3h\,51m of observations, which covered an airmass ranging from 1.97 to 1.16 and were approximately centered on the time of mid-transit given by the ephemeris of \cite{Mocnik2017}.

Our observations were taken using a ABBA nod pattern to improve sky subtraction, meaning that nod pairs needed to be combined during the data reduction, as described in section \ref{sec:reduction}. We used the $0.432 \times 12$ arcsec slit with an exposure time of 300\,s, other than for 8 frames during ingress where an exposure time of 400\,s was used. This was because of a drop in counts, which was caused by the slit wheel inadvertently changing to a narrower, 0.144 arcsec-wide slit for 5 frames. The slit change occurred during an AB nod pair, which was removed from further analysis due to the different resolution of the A and B nods. This left us with 34 spectra of the target (17 nod pairs), with 4 frames taken with the narrower slit (data points 6 and 7 in the light curve, Fig.\ \ref{fig:he_light_curve}).

We acquired observations of a telluric standard A0 star prior to the transit of WASP-107b, but chose to use ESO's Molecfit \citep{molecfit1,molecfit2} to perform the telluric correction, as described in section \ref{sec:reduction}. We also acquired halogen lamp flats and NeArKrXe arcs, which were used in the data reduction as described in section \ref{sec:reduction}.

\section{Data reduction} \label{sec:reduction}

All of the observed data were reduced using the IDL-based REDSPEC software package\footnote{\url{https://www2.keck.hawaii.edu/inst/nirspec/redspec.html}} \citep{McLean2003,McLean2007}. The package performs standard bad pixel interpolation, removal of fringing effects, and flat-fielding, as well as spatial rectification of curved spectra. We focused our reduction on NIRSPEC order 70, which contains the He\,I triplet at 10833\,\AA\ and covers a wavelength range of 10799--11014\,\AA. The spectra were rectified and extracted in differenced nod pairs so that the sky background and OH emission lines were removed. Aperture photometry was performed to extract the spectra, with an aperture width of 11 pixels.

Following the extraction of the spectra, we used iSpec\footnote{https://www.blancocuaresma.com/s/iSpec} \citep{ispec1,ispec2} to perform the continuum normalization of the spectra and to cut the ends of the spectra where the counts dropped significantly, leaving a usable wavelength range of 10800--10975\,\AA. As mentioned in section \ref{sec:obs}, two of our nod pairs were taken with a narrower, 0.144 arcsec slit. At this stage we used iSpec to degrade the resolution of these spectra to a resolution of 25000, in line with the rest of our data.

To wavelength calibrate our data, we began with the Ar, Ne, Kr, and Xe arc lamp lines taken at the start of the observations. However, we found that this resulted in wavelength solutions that deviated from the truth by $\approx 5$\,km\,s$^{-1}$ (less than half NIRSPEC's resolution of 12\,km\,s$^{-1}$), with deviations that differed by a couple of km\,s$^{-1}$ across the order, indicating a distorted wavelength dispersion. 

To correct these effects, we used stellar and telluric atmosphere models to refine the wavelength solution. For the stellar atmosphere model, we used a PHOENIX model \citep{PHOENIX}, which we degraded to the resolution of NIRSPEC ($R = 25000$) using iSpec. The PHOENIX model had an effective temperature of 4400\,K, surface gravity ($\log g$) of 4.5\,cgs and metallicity ([Fe/H]) of 0.0, as close to WASP-107's parameters as possible ($T_{\mathrm{eff}} = 4430$\,K, $\log g = 4.5$, [Fe/H] = +0.02; \citealt{Anderson2017}).

For the telluric model, we used a synthetic telluric spectrum generated for Mauna Kea with 1.0\,mm precipitable water vapor at an airmass of 1.5, which we obtained from the Gemini Observatory's webpages\footnote{http://www.gemini.edu/sciops/telescopes-and-sites/observing-condition-constraints/ir-transmission-spectra\#0.9-2.7um}. This model had a wavelength spacing of $\Delta \lambda$ = 0.2\,\AA, approximately twice that of our NIRSPEC data. We scaled this model to roughly match the amplitude of the telluric features in our observed spectra. This model was only used for wavelength calibration refinement and not for telluric removal, which is described later in this section.

We applied a barycentric velocity correction to each extracted spectrum using the Astropy library \citep{astropy} in Python and applied a systemic velocity and stellar reflex velocity correction to the PHOENIX model which was convolved with the telluric spectrum for each frame. This resulted in a model containing stellar and telluric absorption lines which was used to refine the wavelength correction of each observed spectrum in turn. Since the He\,I triplet is a chromospheric line, it did not appear in the PHOENIX model we used and so the wavelength refinement was not affected by the planet's He\,I absorption.

To perform the wavelength refinement, we split the spectra into eight separate chunks, which were evenly spaced in wavelength, and cross-correlated each of these with the model using iSpec. It was necessary to split the spectra into chunks for cross-correlation due to the distortion of the wavelength solution across the order. This resulted in the radial velocity of each chunk as a function of wavelength, which we fitted with a cubic polynomial to refine the wavelength dispersion and solution.

Then with the wavelength-corrected extracted spectra in the barycentric frame, we used ESO's Molecfit \citep{molecfit1,molecfit2} to perform the telluric correction. We chose not to use the telluric standard star to perform the telluric removal as the telluric standard was only observed prior to our observations of WASP-107. Since our science observations covered a broad range of airmass, we found that our limited standard star spectra did not adequately remove the telluric absorption from our science spectra.

\begin{figure}
\centering
 \includegraphics[scale=0.4]{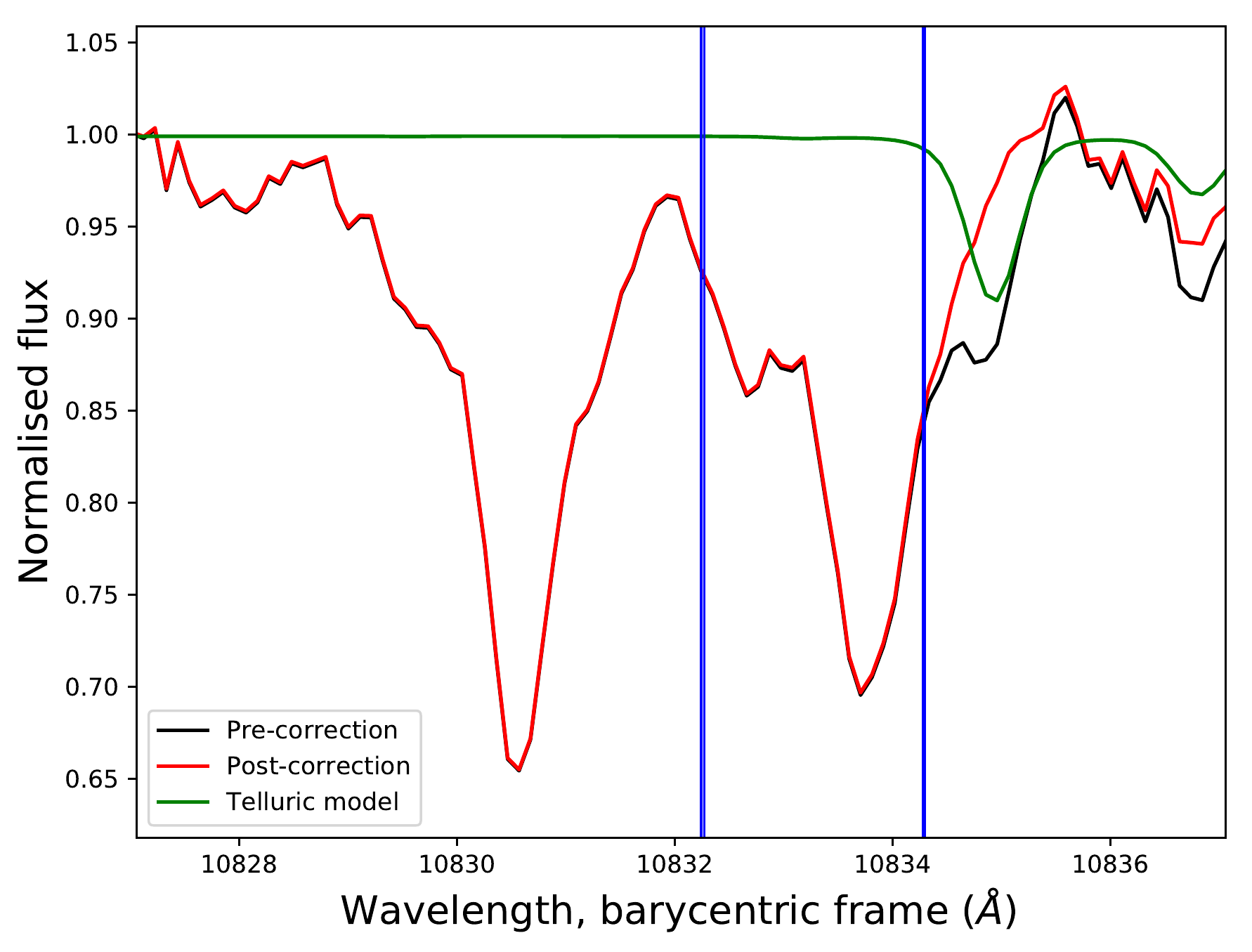}
\caption{The result of the telluric removal performed using Molecfit \protect\citep{molecfit1,molecfit2}. The spectrum before and after the correction is shown in black and red, respectively. The green line shows the telluric absorption model. The location of OH emission lines, which are corrected for by the AB pair subtraction (see text), are shown by the blue vertical lines.}
\label{fig:molecfit_model}
\end{figure}

Molecfit uses Global Data Assimilation System\footnote{https://www.ncdc.noaa.gov/data-access/model-data/model-datasets/global-data-assimilation-system-gdas} (GDAS) profiles which contain weather information for user-specified observatory coordinates, airmasses and times. It then models the telluric absorption lines in the observed spectra using this information. We found that a choice of nine telluric absorption lines, free of significant stellar absorption, allowed Molecfit to perform a good removal of the telluric absorption from our spectra. We chose to fit only for the atmospheric H$_2$O content, with CH$_4$ and O$_2$ fixed. Fig.\ \ref{fig:molecfit_model} shows an example spectrum before and after the telluric correction using Molecfit. This figure demonstrates that while OH emission and telluric absorption lines sit near to the helium triplet lines (which have wavelengths in vacuo of 10832.1, 10833.2 and 10833.3\,\AA), the core of the triplet is unaffected.

With the tellurics removed we then shifted the spectra into the stellar rest frame, using the system parameters given in Table\, \ref{tab:system_params}. We subsequently created master in- and out-of-transit spectra to allow us to study the in-transit absorption by He\,I. Initially we used the ephemeris of \cite{Mocnik2017} to determine which spectra were taken in- and out-of-transit. However, due to the longer duration of the transit at the He\,I triplet, we had to use our own light curve (as discussed in section \ref{sec:discussion}) to determine which phase each spectrum corresponded to. This resulted in three pre/out-of-transit spectra, four ingress spectra, six in-transit spectra, and four egress spectra. We combined our three pre-transit and six in-transit spectra into master out-of-transit and in-transit spectra, respectively.

\begin{table*}
\caption{The system parameters of WASP-107b used in the data reduction and analysis.}
\label{tab:system_params}
\centering

\begin{tabular}{lccc}
Parameter & Symbol & Value & Reference \\ \hline
Time of mid-transit & $T_0$ & 2457584.329746 BJD & \cite{Mocnik2017} \\
Orbital period & $P$ & 5.72149242 d & \cite{Mocnik2017} \\
Orbital inclination & $i$ & 89.560 deg. & \cite{Mocnik2017} \\
White-light scaled planet radius & $R_p/R_*$ & 0.142988 & \cite{Spake2018} \\
Semi-major axis & $a$ & 0.0553 AU & \cite{Mocnik2017} \\
Scaled semi-major axis & $a/R_*$ & 18.10 & \cite{Mocnik2017} \\
Stellar mass & $M_*$ & 0.691\,$\mathrm{M}_{\odot}$ & \cite{Mocnik2017} \\
Planet mass & $M_p$ & 0.119\,$\mathrm{M_J}$ & \cite{Mocnik2017} \\
Planet radius & $R_p$ & 0.924\,$\mathrm{R_J}$ & \cite{Mocnik2017} \\
Semi-amplitude & $K_*$ & 16.45 m\,s$^{-1}$ & \cite{Allart2019} \\
Systemic velocity & $\gamma$ & 13.74\,km\,s$^{-1}$ & Gaia DR2 \citep{Gaia1,Gaia2} \\ \hline

\end{tabular}
\end{table*}

\section{Data analysis} \label{sec:analysis}

\begin{figure}
\centering
\includegraphics[scale=0.45]{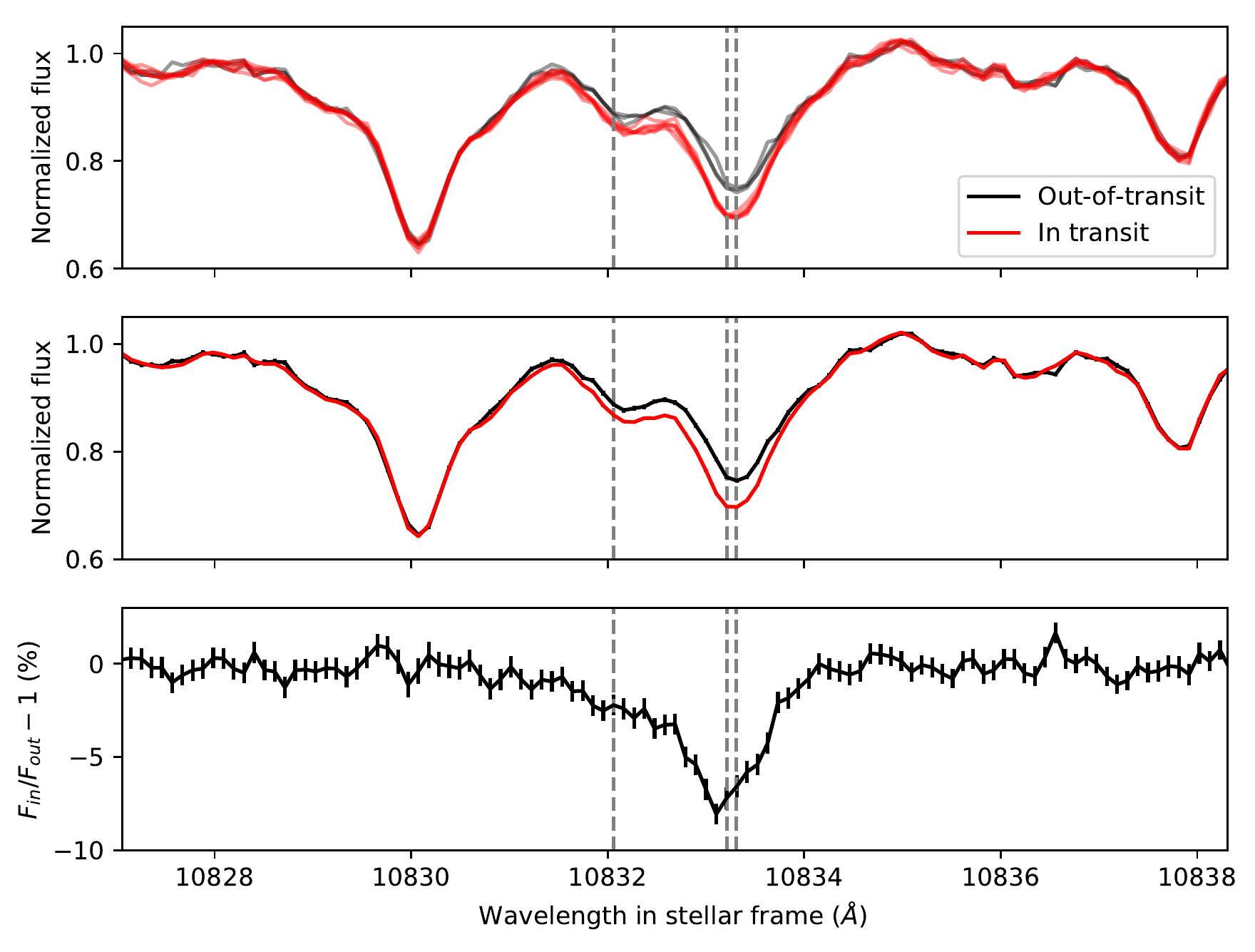}
\caption{Top panel: the individual nod pair spectra of WASP-107. Out-of-transit spectra are shown in black with the in-transit spectra shown in red. This reveals a clear signal of in-transit absorption around the helium triplet, which is shown by the dashed vertical lines. Middle panel: the master co-added nod pair spectra, with out-of-transit shown in black and in-transit spectra shown in red. Bottom panel: the master in-transit spectrum divided by the master out-of-transit spectrum, which reveals the excess absorption by WASP-107b's atmosphere in the region of the helium triplet.}
\label{fig:in_out}
\end{figure}

Fig.\ \ref{fig:in_out} shows the resulting master in- and out-of-transit spectra, in the stellar rest frame. This figure clearly indicates the excess absorption centered on the helium triplet during the planet's transit, which reaches a level of over 7\,\%.

\begin{figure}
\centering
\includegraphics[scale=0.6]{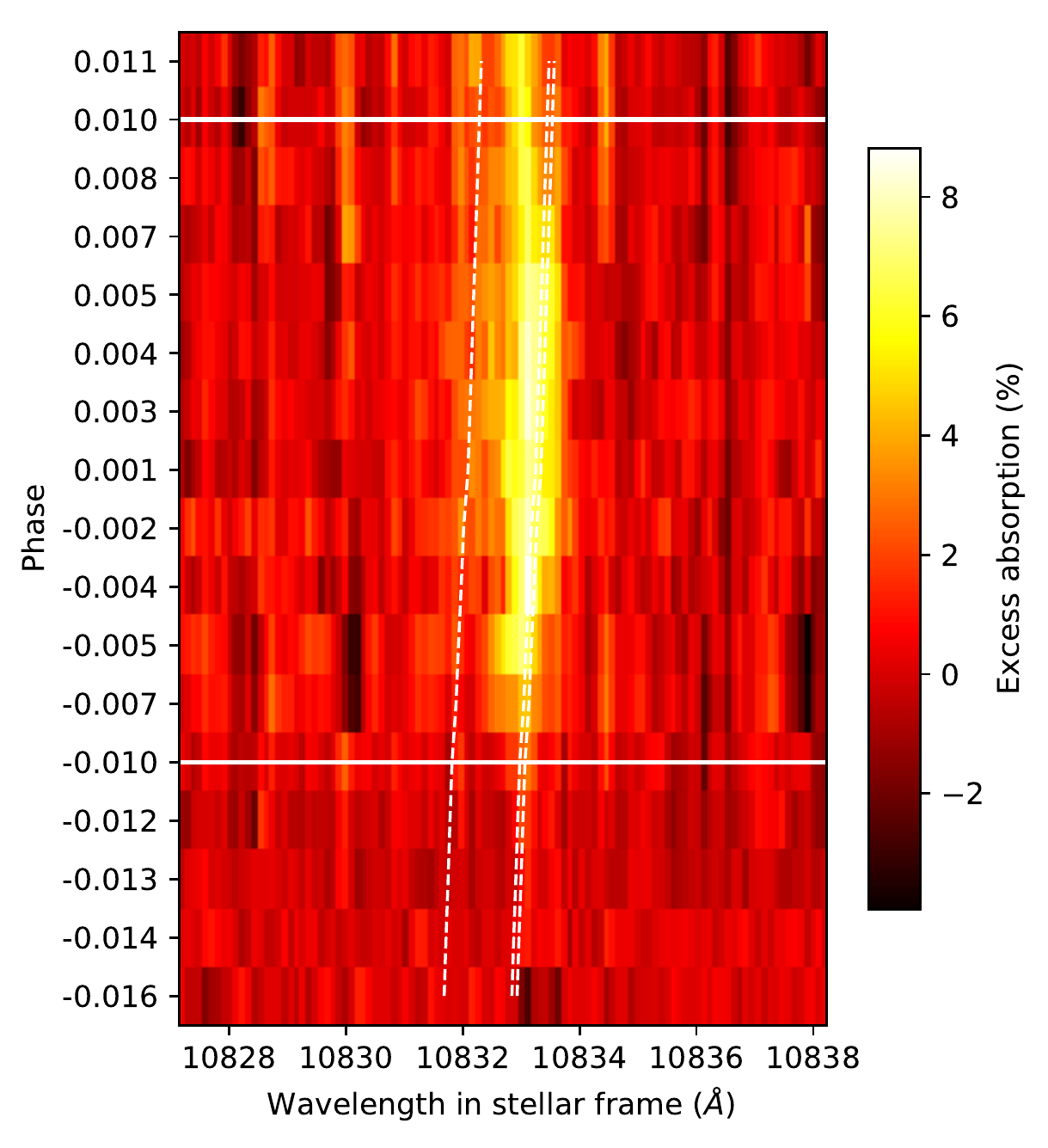}
\caption{The excess absorption by metastable helium during the transit of WASP-107b shown on a plot of orbital phase vs wavelength. The dashed white lines indicate the planet's orbital motion at the wavelengths of the helium triplet. The horizontal white lines indicate the ingress and egress phases from the white light curve of \protect\cite{Spake2018}. Note that the $y$-axis is not evenly spaced in phase.}
\label{fig:heat_map}
\end{figure}

To generate the transmission spectrum and light curve of WASP-107b, we first had to calculate the residual spectra for each frame by dividing each frame by the master out-of-transit spectrum. Fig.\ \ref{fig:heat_map} shows the plot of the excess absorption in the stellar frame as a function of the planet's orbital phase. The dashed white lines indicate the planet's velocity and show that the wavelength of the absorption is consistent with the planetary motion for the majority of the transit but deviates during egress. This could indicate the presence of material trailing the planet, which is being blown away. However, further observations are needed to confirm this feature and hypothesis.

\begin{figure}
\centering
\includegraphics[scale=0.55]{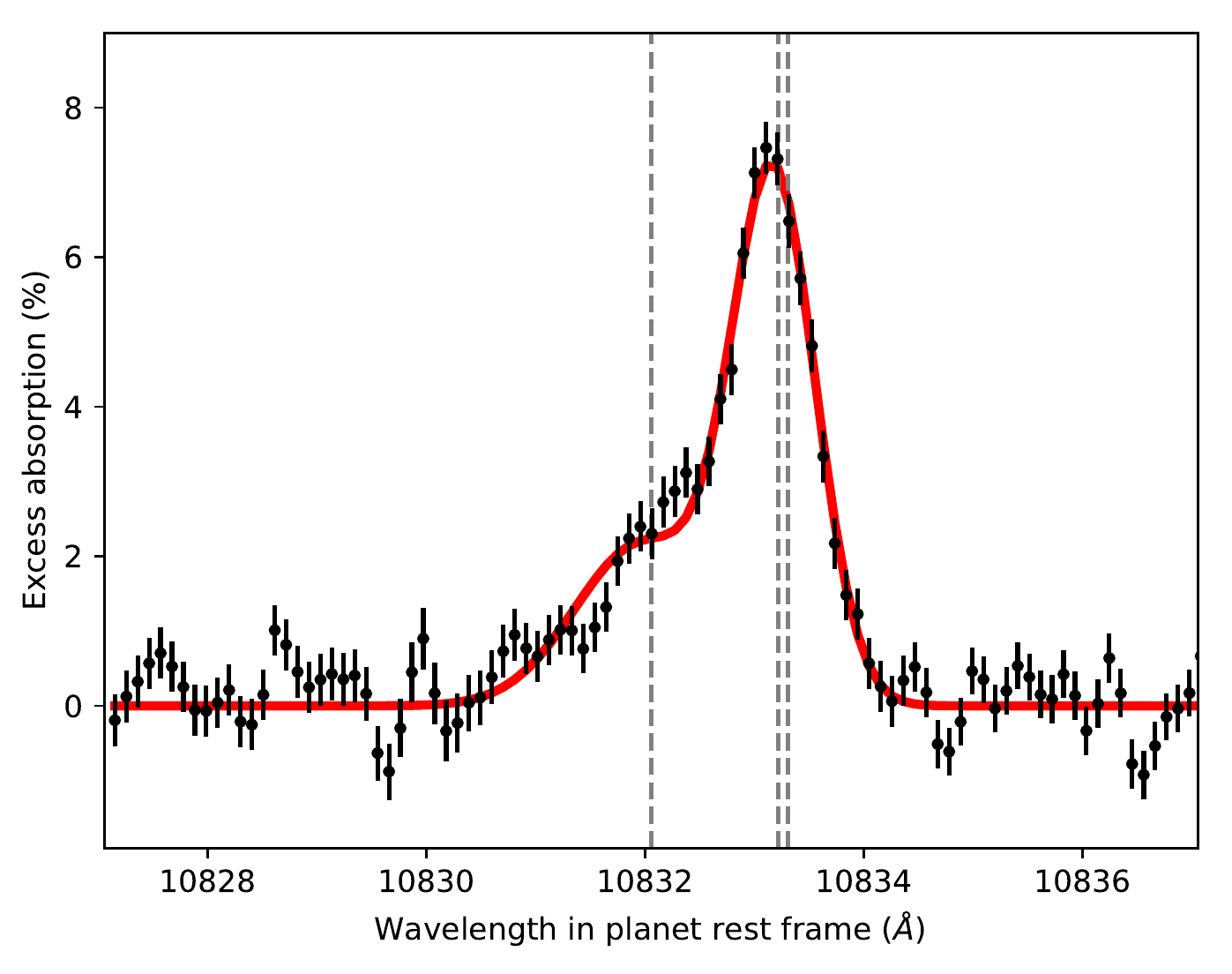}
\caption{The excess absorption during transit, in the planet's rest frame. The vertical dashed lines indicate the locations of the helium triplet and the red line shows a fit of a double-peaked Gaussian to this absorption profile, which resulted in a peak absorption of $7.26 \pm 0.24$\,\% during transit. The centers of the Gaussians are blueshifted by $-2.35 \pm 0.39$\,km\,s$^{-1}$.}
\label{fig:transmission_spectrum}
\end{figure}

\begin{figure}
\centering
\includegraphics[scale=0.48]{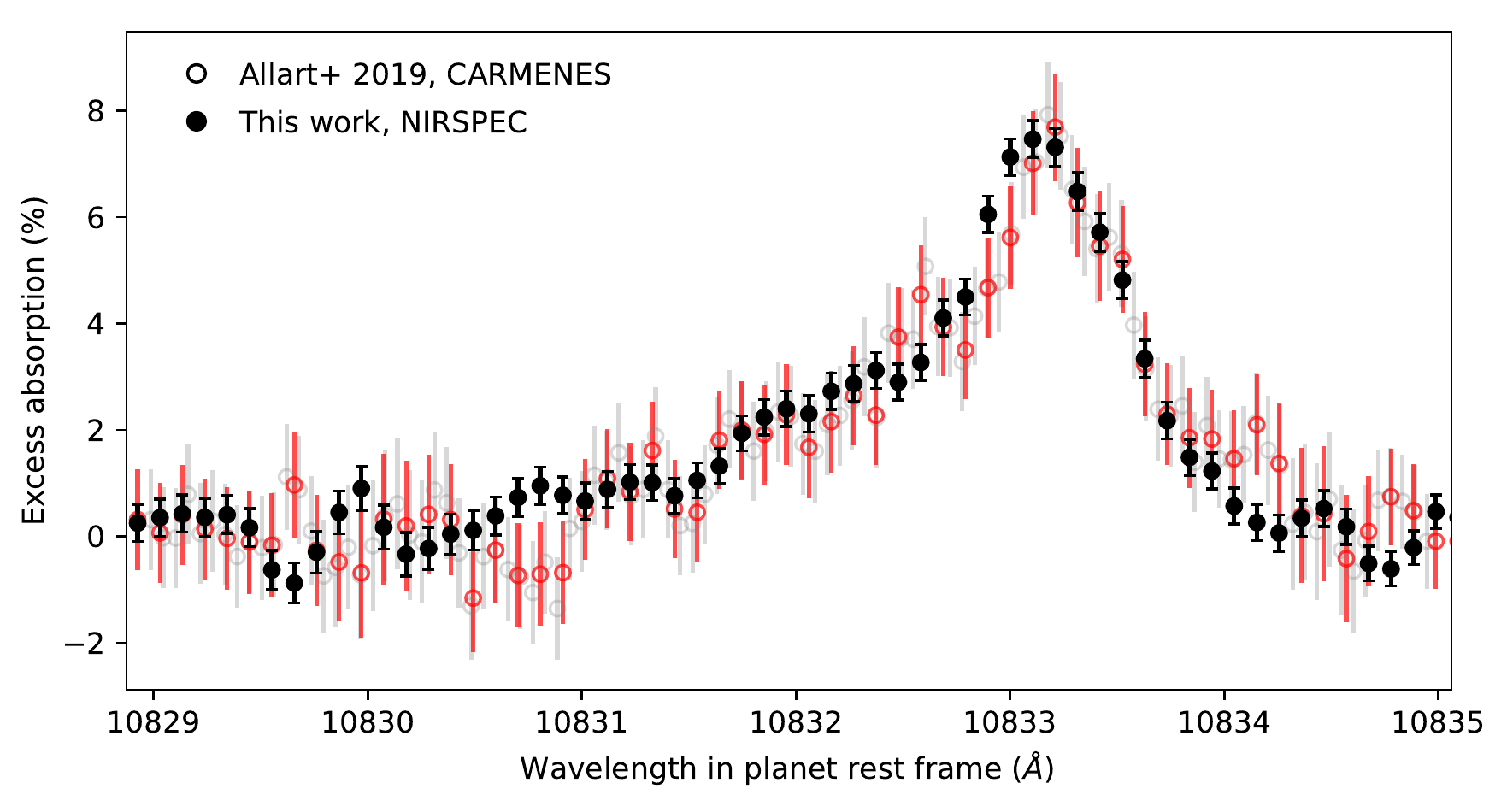}
\caption{The excess absorption during transit, in the planet's rest frame. The results of this work are shown by the filled black circles. The results from \protect\cite{Allart2019}'s study using CARMENES are shown by the open circles, with the grey points at the resolution of CARMENES ($R = 80400$, \protect\citealp{Quirrenbach2018}) and the red points at the resolution of NIRSPEC ($R = 25000$).}
\label{fig:comparison}
\end{figure}

By shifting the excess absorption to the planet's rest frame, we were able to construct the He\,I transmission spectrum of the planet, which is shown in Fig.\ \ref{fig:transmission_spectrum}. This transmission spectrum resulted in a peak excess absorption of $7.26 \pm 0.24$\,\% during transit. We fitted this absorption profile with the summation of two Gaussians using a non-linear least squares fit using the Scipy Python module \citep{scipy}. We fitted for the standard deviations and amplitudes of the two components, along with a wavelength offset which allowed the means of the two Gaussians to deviate from the locations of the absorption wavelengths of the He\,I triplet. This was necessary owing to the small blueshift that was apparent in the peak's location (Fig.\ \ref{fig:transmission_spectrum}), as also noted by \cite{Allart2019}. This resulted in an amplitude for the first and second Gaussian components of $2.17 \pm 0.14$\,\% and $6.95 \pm 0.21$\,\%, respectively. This gave a ratio between the two components of 3.2, which deviates from the optically thin ratio of 8 (e.g.\ \citealp{deJager1966,Salz2018}). The fitted offset was $-0.085 \pm 0.014$\,\AA, or equivalently a blueshift of $-2.35 \pm 0.39$\,km\,s$^{-1}$. This blueshift indicates material is being blown away from the planet, in agreement with the findings of \cite{Allart2019} (Fig.\ \ref{fig:comparison}).

Furthermore, we are able to compare our transmission spectrum with that of \cite{Allart2019} who used CARMENES \citep{Quirrenbach2014} on the 3.5\,m telescope at Calar Alto (Fig.\ \ref{fig:comparison}). This figure demonstrates the excellent agreement between our result and that of \cite{Allart2019}, both in terms of the amplitude and shape of the transmission spectrum. Fig.\ \ref{fig:comparison} also shows that a resolution of 25000 is sufficient to resolve the amplitude of the He\,I absorption. \cite{Allart2019} additionally degraded their spectrum to the resolution of \cite{Spake2018}'s HST/WFC3 spectrum and found excellent agreement with their data. The three studies of WASP-107b's helium atmosphere (\citealp{Spake2018,Allart2019} and this work) therefore indicate that the signal is non-variable over the two year baseline of these studies.

\begin{figure}
\centering
\includegraphics[scale=0.55]{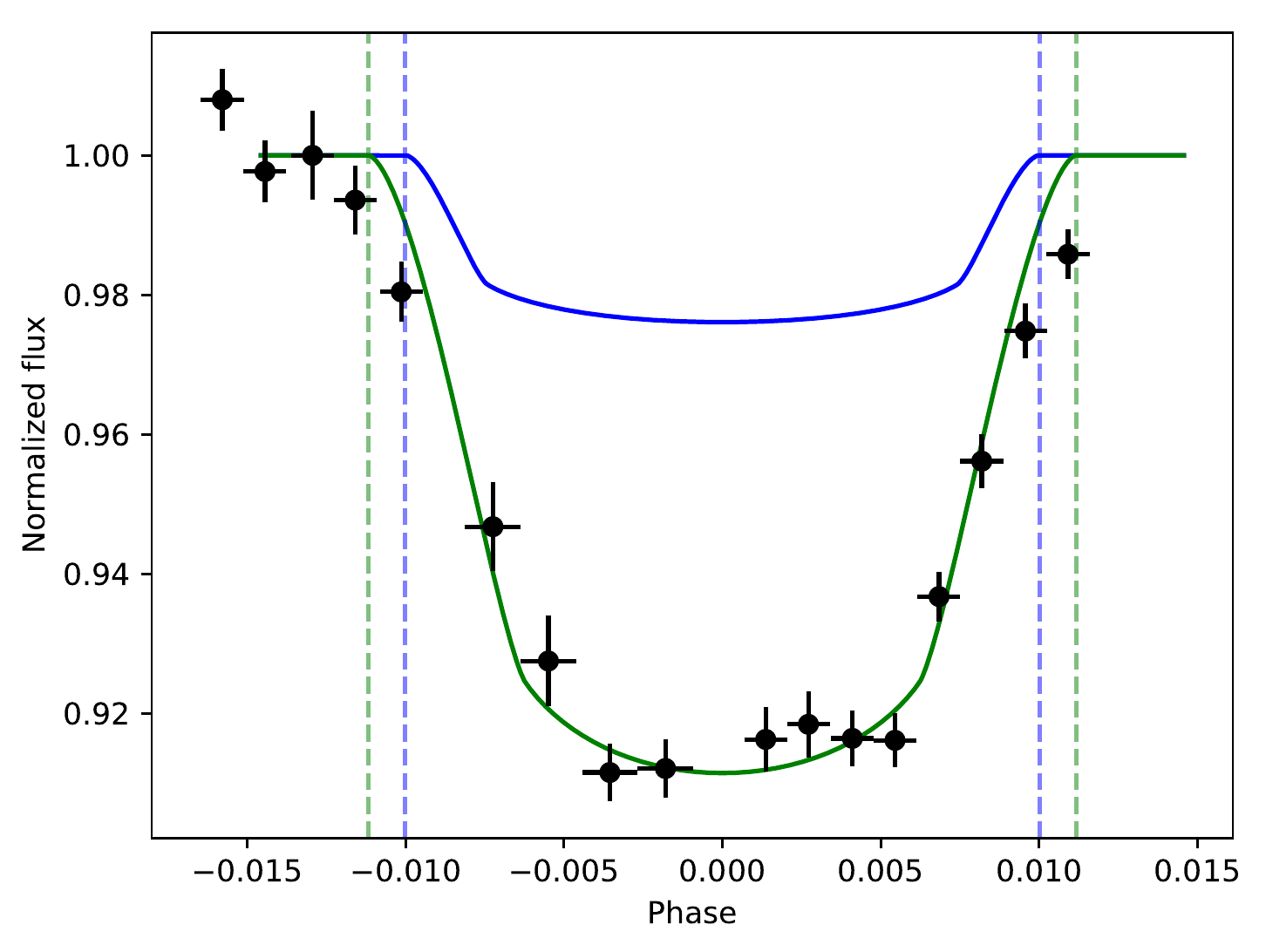}
\caption{The absolute helium light curve integrated from 10833.05 to 10833.48\,\AA\ (black error bars) and following the multiplication by the white light curve of \protect\cite{Spake2018}. The green line shows the transit light curve fitted to our observations, with the blue line showing the white light curve of \protect\cite{Spake2018}. The $R_P/R_*$ of our He light curve is $1.93 \times$ the white light value of \protect\cite{Spake2018}. The vertical dashed blue and green lines indicate the ingress and egress phases of \protect\cite{Spake2018}'s white light curve and our He light curve, respectively. We measure a longer transit duration at the location of the He triplet, with a lower limit on the extra duration of 19 minutes.}
\label{fig:he_light_curve}
\end{figure}

Fig.\ \ref{fig:he_light_curve} shows the absolute light curve constructed by multiplying the excess absorption in the planet's rest frame by a light curve with the same $R_P/R_*$ as the white light curve of \cite{Spake2018}. Our light curve was constructed using a 0.43\,\AA\ region centered on the mean of the redder two lines of the helium triplet (10833.26\,\AA). We fitted this absolute light curve with an analytic transit light curve using the Batman Python module \citep{batman}, using a non-linear limb darkening law with the coefficients fixed to the values used by \cite{Spake2018}. We fixed the inclination ($i$), scaled semi-major axis ($a/R_*$), period ($P$) and transit midpoint ($T_0$) to the values in Table \ref{tab:system_params}. We fitted for $R_P/R_*$ only, which we again did using a non-linear least squares fit through the Scipy Python module \citep{scipy}. This resulted in an $R_P/R_* = 0.2759 \pm 0.0025$, which is $1.93 \times $ the white light $R_P/R_*$ of \cite{Spake2018}.

Fig.\ \ref{fig:he_light_curve} also shows the excess transit duration that we observed at the core of the He triplet as compared with the near-infrared continuum. To determine the extra duration, we resampled our He light curve model and the near-infrared light curve model at a cadence of 30\,s. The difference in the first and fourth contact points between the two models amounted to an excess transit duration of 19 minutes observed in the He line core. This compares to an excess transit duration of 30 minutes observed by \cite{Allart2019} in a 0.75\,\AA -wide bin. However, given our little out-of-transit coverage, our extra duration should be considered a lower limit, while \cite{Allart2019} had better coverage of the transit.

\subsection{Stellar variability}

In this section we consider what effect the intrinsic stellar variability of WASP-107 has on our result (Fig.\ \ref{fig:transmission_spectrum}).

For optical data in particular, stellar activity can lead to deeper or shallower transit depths depending on the temperature of the active region in relation to the photosphere, and whether this activity lies along or away from the planet's transit chord (e.g.\, \citealp{McCullough2014,Oshagh2014,Kirk2016,Rackham2017,Alam2018}).

\cite{Anderson2017} used WASP photometry, acquired in 2009 and 2010, to measure the photometric modulation of WASP-107. They found WASP-107 to modulate with a period of 17 days and an amplitude of 0.4\,\%. \cite{Spake2018} performed a similar analysis using ground-based photometric monitoring of WASP-107 from MEarth \citep{Nutzman2008} and AIT \citep{Henry1999}. From these two instruments, both covering approximately 4 months in 2017, the authors detected modulations of 0.0015 and 0.005 magnitudes with periods of 19.7 and 8.675 days, respectively. \cite{Spake2018} found that a heterogeneous photosphere with a spot covering fraction of $8^{+6}_{-3}$\,\% and faculae covering fraction of $53^{+15}_{-12}$\,\% could lead to the 0.4\,\% variation found by \cite{Anderson2017}.

\cite{Cauley2018} simulated the transit of a planet across an active host star and measured the changes in atomic lines sensitive to the stellar chromosphere, including He\,I at 10833\,\AA. They simulated a number of different scenarios, changing the overall activity level of the star and the latitude of the activity with respect to the planet's transit chord. The maximum spot and faculae\footnote{Note that \cite{Cauley2018} included chromospheric plage in their definition of faculae.} covering fractions they considered were 10 and 50\,\% respectively, similar to the covering fractions for WASP-107 found by \cite{Spake2018}. \cite{Cauley2018} showed that He\,I absorption at 10833\,\AA\ can be contaminated at the 0.1\,\% level (far smaller than the 7.26\,\% absorption we detect, Fig.\ \ref{fig:transmission_spectrum}). Also, depending on the location of this activity, it can actually lead to a dilution of the planet's absorption signal rather than an enhancement.

Furthermore, given that the amplitude of the absorption is consistent over a baseline of 2 years (\citealp{Spake2018,Allart2019}), it is unlikely that stellar activity, which is inherently variable, could produce such consistency. The above reasons indicate that the absorption we detect (Fig.\ \ref{fig:transmission_spectrum}) is planetary in nature and is not significantly influenced by stellar activity.

\subsubsection{The Rossiter-McLaughlin effect}

The Rossiter-McLaughlin \citep{Rossiter1924,McLaughlin1924} effect can lead to spurious velocity shifts if not correctly accounted for in the planet's frame. However, it has less of an effect in the stellar frame where the Rossiter-McLaughlin effect is near-symmetric (e.g.\ \citealp{Louden2015}). In the case of our observations, the small blueshifted absorption we observe is present in both the stellar rest frame (Fig.\ \ref{fig:in_out}) and the planet's rest frame (Fig.\ \ref{fig:transmission_spectrum}).

Since WASP-107b is a slow rotator ($v \sin i = 2.5 \pm 0.8$\,km\,s$^{-1}$, \citealp{Anderson2017}), the predicted amplitude of the Rossiter-McLaughlin effect is expected to be small \citep{Dai2017}. Indeed, using equation 5 of \cite{Dai2017}, with the $R_P/R_*$ we calculate from our He light curve in section \ref{sec:analysis}, we find that the expected signal amplitude is 133\,m\,s$^{-1}$. This is an order of magnitude smaller than the blueshift we observe (Fig.\ \ref{fig:transmission_spectrum}) and two orders of magnitude smaller than NIRSPEC's resolution. We therefore do not expect our results to be impacted by the Rossiter-McLaughlin effect.

\section{Discussion}
\label{sec:discussion}

\subsection{WASP-107b's extended helium atmosphere}

This is the third paper that presents the detection of the extended helium atmosphere of WASP-107b, confirming the results of \cite{Spake2018} and \cite{Allart2019}. We find that the planet's radius is $1.93 \times$ larger at the location of the helium triplet than the surrounding continuum (section \ref{sec:analysis}, Figs.\ \ref{fig:transmission_spectrum} and \ref{fig:he_light_curve}). This amounts to approximately half the Roche lobe radius of 3.34\,R$_{\mathrm{P}}$, using the approximation of \cite{Eggleton1983} and planet parameters given in Table \ref{tab:system_params}. 

The helium absorption profile of \cite{Allart2019} showed a blueshifted excess, indicating a tail of escaping material. They modeled this absorption using the 3D EVaporating Exoplanet code (EVE) \citep{Bourrier2013,Bourrier2016} and found that their simulations were consistent with helium escaping at the exobase with a thermal wind velocity of $\sim 12$\,km\,s$^{-1}$. 

The shape of the excess absorption by He\,I we detect is very similar to that seen by \cite{Allart2019} (Fig.\ \ref{fig:comparison}). Similar to that study, we also detect a blueshifted excess (Fig.\ \ref{fig:transmission_spectrum}), which we measure to have an amplitude of $-2.35 \pm 0.39$\,km\,s$^{-1}$. This is further evidence for a wind of material escaping the planet, following the conclusions of \cite{Spake2018} and \cite{Allart2019}. We also detect non-Keplerian, blueshifted absorption during WASP-107b's egress (Fig.\ \ref{fig:heat_map}), which could correspond to material trailing the planet. However, additional observations are needed to confirm this feature.

Given the absence of any post-transit, and only a short pre-transit, baseline we can only place a lower limit on the excess transit duration at the location of the helium triplet (Fig.\ \ref{fig:he_light_curve}). The excess transit duration we observe is 19 minutes, as compared with the white light transit of \cite{Spake2018}. Our He\,I light curve also appears symmetric about the mid-point (Fig.\ \ref{fig:he_light_curve}), similar to what was found by \cite{Spake2018} and \cite{Allart2019}. However, we note that our lack of post-transit baseline does not allow us to constrain the presence of post-transit absorption.

\subsection{Keck/NIRSPEC as an instrument for exoplanetary He\,I observations}


Our study demonstrates Keck/NIRSPEC's ability to detect the helium triplet in an exoplanet's atmosphere, making it the third ground-based high resolution spectrograph to successfully detect this feature. This follows the use of CARMENES \citep{Quirrenbach2014} and the Habitable Zone Planet Finder (HPF; \citealt{HPF1,HPF2}) near-infrared spectrograph on the 10 meter Hobby-Eberly Telescope (HET), which have together made similar detections in six other exoplanets \citep{Allart2018,Allart2019,Alonso-Floriano2019,Ninan2019,Nortmann2018,Salz2018}.

The resolutions of CARMENES ($R \approx 80400$) and HPF ($R \approx 55000$) are both higher than NIRSPEC ($R \approx 25000$). This offers advantages with respect to detecting the blueshifted absorption profile expected to be associated with an escaping wind of material. However, Fig.\ \ref{fig:comparison} demonstrates that a resolution of 25000 is sufficient to resolve the absorption's amplitude and shape for WASP-107b. Additionally, Keck II's 10 meter aperture does provide a significant advantage in terms of S/N over the 3.5\,m telescope at Calar Alto Observatory that is home to CARMENES. This is highlighted in Fig.\ \ref{fig:comparison}, which shows the higher precision that we were able to achieve with our Keck observations, as compared to \cite{Allart2019}. Our demonstration is promising for the search for helium signatures around smaller planets, where the higher precision will be more important than resolution in the search for these smaller signals.

\subsection{Predicting promising exoplanets for observations of He\,I}

In addition to WASP-107b (\citealp{Spake2018,Allart2019}, this work) extended helium atmospheres have been detected for the hot Jupiters HD\,189733b \citep{Salz2018} and HD\,209458b \citep{Alonso-Floriano2019}, the Saturn-mass planet WASP-69b \citep{Nortmann2018}, and the Neptunes HAT-P-11b \citep{Allart2018,Mansfield2018} and GJ\,3470b \citep{Ninan2019}.  Four of these exoplanets orbit K stars while the remaining two orbit G and M stars. \cite{Oklopcic2019} suggested that K stars have the necessary XUV (EUV and X-ray) to mid-UV flux ratios to maintain a populated metastable helium state in the atmospheres of these exoplanets, as EUV ionizes the helium ground state, populating the metastable state, while mid-UV ionizes the helium metastable state. This makes K-star hosts the most favorable targets for detecting exoplanetary metastable helium. Similarly, \cite{Nortmann2018} showed that the exoplanets with helium detections orbit stars with higher activity levels and receive greater levels of XUV radiation than the exoplanets with non-detections. 
 
In addition to the host's spectral type, both the gravitational potential and semi-major axis of the planet are predicted to influence the atmospheric escape \citep{Salz2016a,Nortmann2018,Oklopcic2019}. 

In Table \ref{tab:he_sample} we present 11 known exoplanets that are promising targets for future observations of extended helium atmospheres. To derive this list, we took the sample of well-studied transiting planets from TEPCat\footnote{https://www.astro.keele.ac.uk/jkt/tepcat/} \citep{Southworth2011} and selected planets that have bloated radii, which lead to large transit depths per scale height ($\delta$) of $> 100$\,ppm, calculated through

\begin{equation}
\delta = \frac{2 H R_p}{R_*^2},
\end{equation}

where $H$ is the atmospheric scale height, and $R_P$ and $R_*$ are the radii of the planet and star, respectively. We then kept only those planets that orbit relatively bright ($J \leq 11$) K stars ($4000 \leq \mathrm{T_{eff}} \leq 5100$\,K). 

We also note that the gravitational potentials of 10 of these planets fall within the hydrodynamic wind regime of \cite{Salz2016a} ($\log_{10} (- \Phi_{\mathrm{G}}) < 13.2$), while Qatar-1b sits in an intermediate region where hydrodynamic escape can exist but is suppressed \citep{Salz2016b}.  

In Table \ref{tab:he_sample} we also include the semi-major axes of this sample, as \cite{Nortmann2018} and \cite{Oklopcic2019} showed that helium absorption depends on the orbital separation. \cite{Oklopcic2019} suggested that semi-major axes between $\sim 0.03$ and 0.1\,AU are optimal for observations of helium in exoplanets orbiting main sequence stars.

\begin{figure*}
\centering
\includegraphics[scale=0.48]{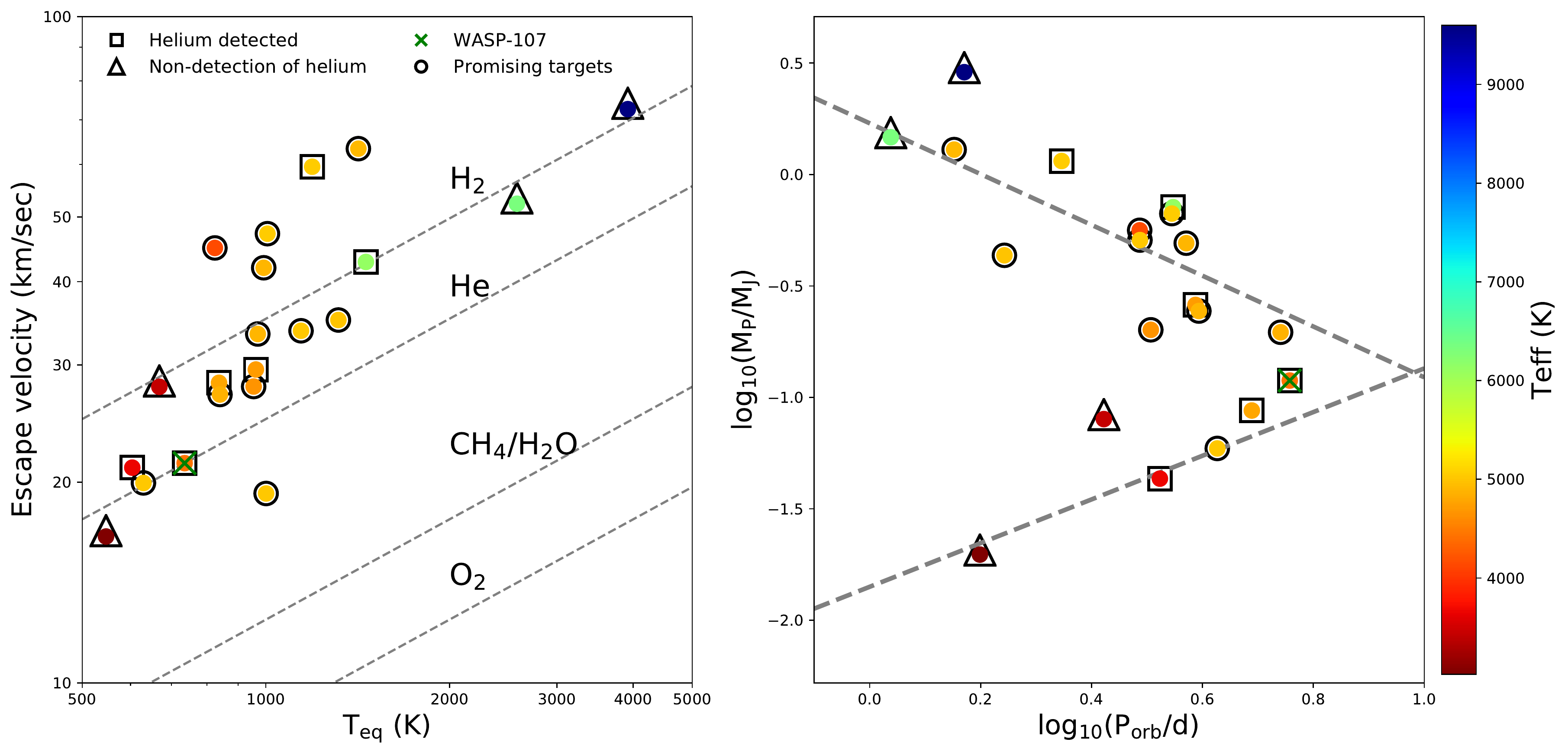}
\caption{Left panel: a plot of planet escape velocity against equilibrium temperature for exoplanets with confirmed He\,I detections (black squares), non-detections (black triangles), and the sample of exoplanets that we have identified as promising targets for helium observations (black circles). WASP-107b is highlighted by the green cross. All the plotted exoplanets are colored by the effective temperatures of their host stars. The grey dashed lines indicate the threshold escape limits for various gas compositions considering thermal escape (Jeans escape). Planets falling below a given line are expected to be losing that gas. Note: the temperature in the escaping zone (upper atmosphere) of the envelope may be higher than the estimated equilibrium temperature of the planet, due to absorption of UV photons by various gas species and UV-driven photochemistry. However, the equilibrium temperature can be considered a lower bound. Right panel: the same sample of planets as shown in the left panel but shown in mass-period space. The dashed grey lines indicate the extent of the Neptune desert, as defined by \protect\cite{Mazeh2016}.}
\label{fig:he_sample}
\end{figure*}

Fig.\ \ref{fig:he_sample} (left panel) plots the escape velocity of this sample of planets against their equilibrium temperatures, with the mean velocity of various gases overplotted. Any planet sitting below a certain line is susceptible to losing that gas via Jeans escape. This demonstrates that hydrodynamic escape is dominating for the sample of planets with detected helium absorption, as they all sit above the line where helium would be lost via Jeans escape. 

We also plot this sample of planets in mass-period space (Fig.\ \ref{fig:he_sample}, right panel), along with the boundaries of the Neptune desert as defined by \cite{Mazeh2016}. The 11 exoplanets in Table \ref{tab:he_sample} sample this space well, and observations of He\,I could add further constraints on how this desert formed, whether it be via atmospheric loss driven by stellar XUV radiation (e.g.\ \citealp{Allan2019}), Roche-lobe overflow (e.g.\ \citealp{Matsakos2016}), or a combination of the two (e.g.\ \citealp{Kurokawa2014}). A better understanding of the processes shaping the Neptune desert might help with the interpretation of the radius-gap between 1.5 and 2 Earth radii \citep{Fulton2017}.

\begin{table*}
\caption{The sample of 11 planets that we predict to show extended helium atmospheres.}
\label{tab:he_sample}
\centering
\begin{tabular}{lcccccccc}
 &  $\mathrm{T_{eff}}$ &     $\mathrm{T_{eq}}$ &  Escape vel. & $\log_{10} (- \Phi_{\mathrm{G}})$ & Semi-major &  J mag.$^b$ &  Transit depth  & Reference \\ 
 & (K) & (K) &  (km\,s$^{-1}$)$^a$ & (erg\,g$^{-1}$)$^a$ & axis (AU) &  &  per $H$ (ppm)$^a$ &  \\ \hline
 HAT-P-12b &  4665 &   955 &        27.8 & 12.59  & 0.03767 & 10.8 &      342 & \cite{Mancini2018} \\
 HAT-P-18b &  4870 &   841 &        27.1 & 12.56 & 0.0559 & 10.8 &      303 &  \cite{Esposito2014}  \\
 HAT-P-26b &  5011 &  1001 &        19.2 & 12.27 & 0.0479 & 10.1 &      211 & \cite{Hartman2011} \\
 Qatar-1b  &  4910 &  1418 &        63.3 & 13.30 & 0.02332 & 11.0 &      109 & \cite{Collins2017} \\
 Qatar-6b  &  5052 &  1006 &        47.2 & 13.05 & 0.0423 & 9.7 &      148 & \cite{Alsubai2018} \\
 TOI-216b &  5026 &   628 &        20.0 & 12.30 & 0.1293 & 10.8 &      182 & \cite{Kipping2019} \\ 
 WASP-11b &  4900 &   992 &        42.0 & 12.94 & 0.04375 & 10.0 &      140 & \cite{Mancini2015} \\
 WASP-29b &  4875 &   970 &        33.4 &  12.75 &  0.04565 &  9.4 &      122 & \cite{Gibson2013} \\
 WASP-52b &  5000 &  1315 &        35.0 & 12.79 & 0.02643 & 10.6 &      413 & \cite{Mancini2017} \\ 
 WASP-80b &  4145 &   825 &        44.9 & 13.00 & 0.03479 & 9.2 &      171 & \cite{Mancini2014} \\
 WASP-177b &  5017 &  1142 &        33.8 & 12.76 & 0.03957 & 10.7 &      484 & \cite{Turner2019} \\ \hline

\multicolumn{9}{l}{$^a$derived parameter using values from studies in the reference column.} \\
\multicolumn{9}{l}{$^b$2MASS magnitudes \protect\citep{Skrutskie2006}.} \\

\end{tabular}

\end{table*}

\section{Conclusions}
\label{sec:conclusions}

We have detected He\,I at 10833\,\AA\ in the extended atmosphere of WASP-107b, using Keck II/NIRSPEC. The excess absorption we detect has a peak amplitude of $7.26 \pm 0.24$\,\% and is blueshifted by $-2.35 \pm 0.39$\,km\,s$^{-1}$. We also see evidence for non-Keplerian, blueshifted absorption during the planet's egress, which could be the result of material trailing the planet. This is further confirmation of WASP-107b's extended helium atmosphere, which is being actively lost, following the findings of \cite{Spake2018} and \cite{Allart2019}.

The amplitude and shape of the helium absorption profile we detect are in excellent agreement with the CARMENES results of \cite{Allart2019}, who in turn demonstrated consistency with \cite{Spake2018}'s original HST detection. Our result, when taken together with those of \cite{Spake2018} and \cite{Allart2019}, demonstrates that the helium absorption of WASP-107b does not show temporal variability over the baseline of these three studies. We measure an $R_P/R_*$ that is $1.93 \times$ the white light value of \cite{Spake2018} and a transit duration that is a minimum of 19 minutes longer. We are unable to put a strong constraint on the extra transit duration owing to our short out-of-transit baseline.

This is the first time Keck II/NIRSPEC has been used to detect helium in an exoplanet atmosphere, demonstrating its capability to probe the upper atmospheres of highly irradiated exoplanets. NIRSPEC now joins CARMENES and HPF as ground-based high resolution spectrographs that have detected exoplanetary helium at 10833\,\AA. The seven ground-based detections of this line demonstrate its accessibility to such instruments, while the rate of these detections offers exciting prospects for understanding exoplanet evaporation, and its role in carving such features as the Neptune desert (e.g.\ \citealp{Mazeh2016}) and radius-gap \citep{Fulton2017}, across a large sample of planets.

\section{Software and third party data repository citations} \label{sec:cite}

%

%
%


\acknowledgments

We are grateful to the anonymous referee whose comments led to a significant improvement in the manuscript. We are also grateful to Romain Allart for sharing the CARMENES transmission spectrum of WASP-107b, and to Antonija Oklop{\v{c}}i{\'c} for useful discussions regarding NIRSPEC and the interpretation. Finally we also wish to thank our support astronomer, Greg Doppmann, for guidance during the reduction stage.

The data presented herein were obtained at the W. M. Keck Observatory, which is operated as a scientific partnership among the California Institute of Technology, the University of California and the National Aeronautics and Space Administration. The Observatory was made possible by the generous financial support of the W. M. Keck Foundation. The authors wish to recognize and acknowledge the very significant cultural role and reverence that the summit of Maunakea has always had within the indigenous Hawaiian community.  We are most fortunate to have the opportunity to conduct observations from this mountain.

%

\vspace{5mm}
\facilities{Keck(NIRSPEC)}


\software{Astropy \citep{astropy}, Batman \citep{batman}, iSpec \citep{ispec1,ispec2}, Matplotlib \citep{matplotlib}, Molecfit \citep{molecfit1,molecfit2}, Numpy \citep{numpy}, REDSPEC \citep{McLean2003,McLean2007}, Scipy \citep{scipy}}

\bibliography{wasp107_bib}

%
%
%



\end{document}